\documentclass[manuscript]{emulateapj}
\usepackage{natbib}
\usepackage{amsmath}
\usepackage{amssymb}
\usepackage{soul}
\usepackage{xcolor}
\DeclareMathSymbol{\circledstar}   {\mathbin}{AMSa}{"46}

\bibpunct{(}{)}{;}{a}{}{,}
\usepackage{txfonts}

\def\bbf#1{{\color{blue}{#1}}}  

\def\st#1{} 
\def\bbf#1{{#1}} 

\begin{document}
%
\title{Deep wideband single pointings and mosaics in radio interferometry - 
How accurately do we reconstruct 
intensities and spectral indices of faint sources? }

\shorttitle{Wide band imaging accuracy} 
\shortauthors{Rau, Bhatnagar \& Owen}


\author{U.~Rau}
\affil{National Radio Astronomy Observatory, Socorro, NM - 87801, U.S.A.}
\email{rurvashi@nrao.edu}
\author{S.~Bhatnagar}
\affil{National Radio Astronomy Observatory, Socorro, NM - 87801, U.S.A.}
\author{F.~N.~Owen}
\affil{National Radio Astronomy Observatory, Socorro, NM - 87801, U.S.A.}

\thanks{The National Radio Astronomy Observatory is a facility of the
  National Science Foundation operated under cooperative agreement by
  Associated Universities, Inc.}

\date{Received: 04/09/2016;  Accepted:05/19/2016 }

\begin{abstract}
  Many deep wide-band wide-field radio interferometric surveys are
  being designed to accurately measure intensities, spectral indices
  and polarization properties of faint source populations.  In this
  paper we compare various wideband imaging methods to evaluate the
  accuracy to which intensities and spectral indices of sources close to
  the confusion limit can be reconstructed.  We simulated a wideband
  single-pointing (C-array, L-Band (1-2GHz)) and 46-pointing mosaic
  (D-array, C-Band (4-8GHz)) JVLA observation using realistic
  brightness distribution ranging from $1\mu$Jy to $100m$Jy and time-,
  frequency-, polarization- and direction-dependent instrumental
  effects.  The main results from these comparisons are (a) errors in
  the reconstructed intensities and spectral indices are larger for
  weaker sources even in the absence of simulated noise, (b) errors
  are systematically lower for joint reconstruction methods (such as
  MT-MFS) along with A-Projection for accurate primary beam
  correction, and (c) use of MT-MFS for image reconstruction
  eliminates Clean-bias (which is present otherwise).  Auxiliary tests
  include solutions for deficiencies of data partitioning methods
  (e.g. the use of masks to remove clean bias and hybrid methods to
  remove sidelobes from sources left undeconvolved), 
  the effect of sources not at pixel centers and the
  consequences of various other numerical approximations within
  software implementations.  This paper also demonstrates the level of
  detail at which such simulations must be done in order to reflect
  reality, enable one to systematically identify specific reasons for
  every trend that is observed and to estimate scientifically
  defensible imaging performance metrics and the associated
  computational complexity of the algorithms/analysis procedures.

\end{abstract}


\maketitle
%

\section{Introduction}\label{s:intro}

The recent upgrade of the Very Large Array (VLA) has resulted in a greatly increased
imaging sensitivity due to the availability of large instantaneous bandwidths at the
receivers and correlator. At least two new dish array telescopes 
(in particular, ASKAP and MeerKAT) 
are currently under construction to improve upon the VLA's specifications in terms of 
instantaneous sky coverage and total collecting area. 
A considerable amount of observing time has been allotted on all three instruments
for large survey projects that need deep and sometimes high dynamic range imaging 
over fields of view that span one or more primary beams. 
Desired data products include images and high precision catalogs of
source intensity, spectral index, polarized intensity and rotation measure, produced 
by largely automated imaging pipelines.
For these experiments, data sizes range from a few hundred 
Gigabytes up to a few Terabytes and contain a large number of frequency channels 
for one or more pointings. 

In this imaging regime, traditional algorithms have limits in the achievable dynamic range
and accuracy with which weak sources are reconstructed. 
Narrow-band approximations of the sky brightness 
and instrumental effects result in sub-optimal continuum sensitivity
and angular resolution. Narrow-field approximations that ignore the time-, frequency-, and
polarization dependence of antenna primary beams prevent accurate reconstructions 
over fields of view larger than the inner part of the primary beam. Mosaics constructed by
stitching together images reconstructed separately from each pointing often have a lower
imaging fidelity than a joint reconstruction. Despite these drawbacks, there are 
several science cases for which such accuracies will suffice. 
Further, all these methods are easy to apply using readily available and stable software and 
are therefore used regularly. 

More recently-developed algorithms that address the above shortcomings also exist.
Wide-field imaging algorithms
\citep{WPROJECTION,AWProjection}
include corrections for instrumental effects such as the w-term and antenna 
aperture illumination functions. 
Wide-band imaging algorithms such as 
Multi-Term Multi-Frequency-Synthesis (MT-MFS) \citep{MSMFS, MFCLEAN} make use of the
combined multi-frequency spatial frequency coverage while reconstructing both the
sky intensity and spectrum at the same time. 
Wideband A-Projection \citep{WBAWP}, a 
combination of the two methods mentioned above 
accounts for the frequency dependence of the sky separately from that of the 
instrument during wideband imaging. Algorithms for joint mosaic reconstruction
\citep{Cornwell88} add together data from multiple pointings either in the spatial-frequency
or image domain and take advantage of the combined spatial-frequency coverage
during deconvolution. Such joint mosaic imaging along with a wideband sky model and 
wideband primary beam correction 
has recently been demonstrated to work accurately and is currently being commissioned
\citep{WBMOS2014}(in prep).
These methods provide superior numerical results compared to traditional methods
but they require all the data to be treated together during the reconstruction and 
need specialized software implementations that are optimized for the large amount
of data transport and memory usage involved in each imaging run.

With so many methods to choose from and various trade-offs between numerical 
accuracy, computational complexity and ease of use, it becomes important to identify
the most appropriate approach for a given imaging goal and to quantify the errors
that would occur if other methods are used.
The Square Kilometre Array (SKA) will involve much larger
datasets than the VLA, ASKAP or MeerKAT will encounter with even more 
stringent accuracy requirements, making it all the more
relevant to characterize all our algorithmic options and use existing, smaller 
instruments to derive and validate algorithmic parameters.
This paper describes some preliminary results based on a series of simulated tests
of deep wide-band and wide-field mosaic observations with the VLA. 

Section \ref{Sec:sims} describes how the datasets were simulated.
Sections \ref{Sec:algos:single1}--\ref{Sec:algos:mosaic} list the
imaging methods that were compared, for the single pointing as well as
the mosaic tests.  Section \ref{Sec:metrics} describes the metrics
used to quantify imaging quality.  Sections \ref{Sec:results:single}
and \ref{Sec:results:mosaic} describe the results from several tests
performed with the single-pointing and mosaic datasets.  Section
\ref{Sec:discussion} summarizes the results, discusses what one can
and cannot conclude from such tests, and lists several other tests
that are required before SKA-level algorithmic accuracy predictions
can be made.

\section{Data Simulation}\label{Sec:sims}

\begin{figure}
\includegraphics[width=3.2in]{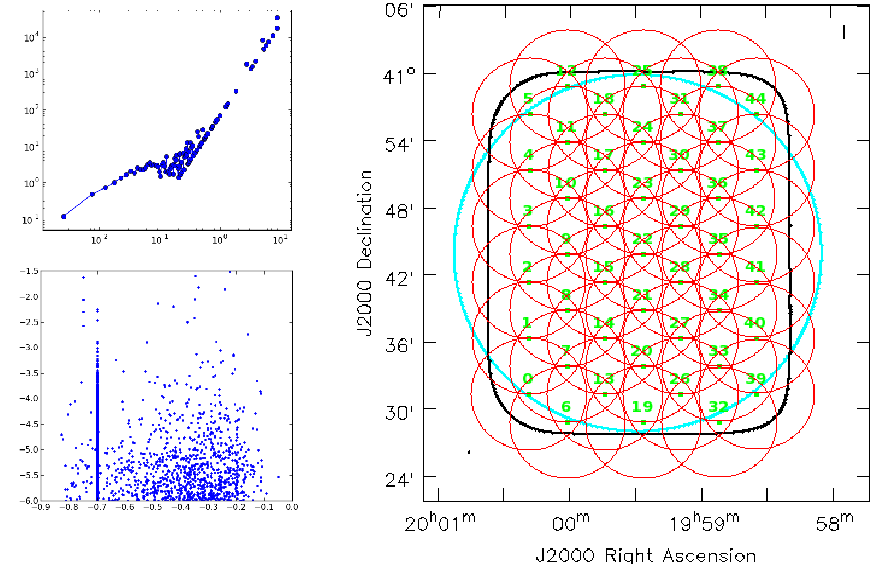}
\caption{Simulated Sky and observations: The top left panel shows the source count vs
intensity for 8000 point sources. The bottom left panel shows intensity vs spectral index
of these sources. The panel on the right shows the sky coverage of the two simulated 
observations. The blue circle represents the half-power point of a single pointing primary beam
at the VLA L-Band. The red circles represent the individual pointings at C-Band and the black
outline shows the half-power level of the mosaic primary beam. }
\label{fig:scounts}
\end{figure}

\begin{figure}
\includegraphics[width=3.2in]{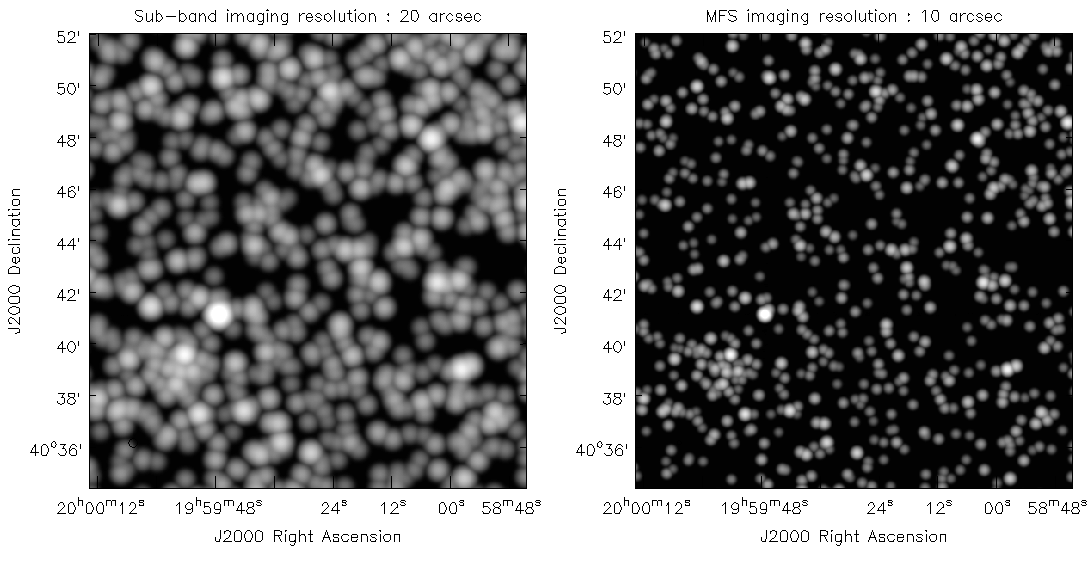}
\includegraphics[width=3.2in]{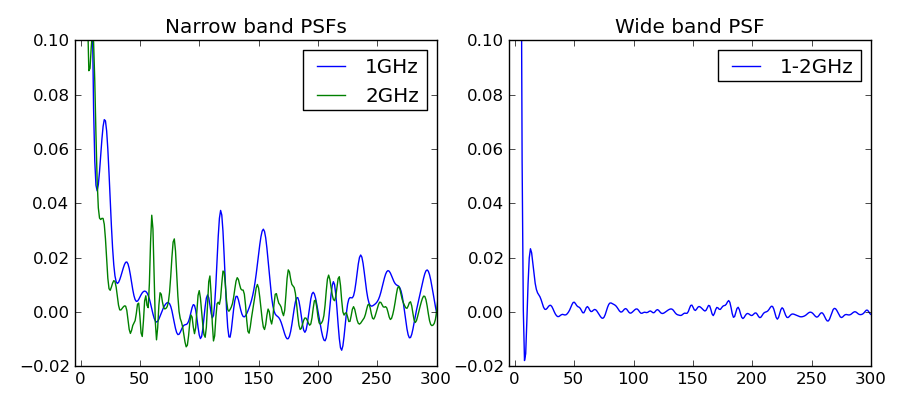}
\caption{Simulated Sky at imaging resolution / PSF sidelobes: 
A small section of the simulated sky is shown in the top panels, smoothed
to 20arcsec (the angular resolution at the low frequency end of the band) in the left panel 
and 10arcsec (corresponding to the MFS angular resolution) in the right panel. 
The lower panels show profiles of the inner sidelobes of the PSFs seen by subband and MFS 
imaging runs.  These images and plots 
illustrate the level of confusion that subband versus MFS based deconvolution algorithms will 
encounter from the main lobe the PSF as well as its inner sidelobes. }
\label{fig:smoothed.model}
\end{figure}

A sky model was chosen to contain a set of 8000 point sources spanning
one square degree in area. The source list is a subset of that
available from the SKADS/SCubed simulated sky project \citep{SKADS}.
In this sample, intensities ranged between $1 \mu Jy$ and $7 mJy$ and
followed a realistic source count distribution.  For high dynamic
range tests, one $100mJy$ source was also added.  Spectral indices
ranged between 0.0 and -0.8 with a peak in the spectral index
distribution at -0.7 plus a roughly Gaussian distribution around -0.3
with a width of 0.5. Fig.~\ref{fig:scounts} shows the source count vs
intensity on the top-left panel and intensity vs spectral index on the
bottom-left\footnote{Such a sharply peaked spectral index
  distribution is not consistent with observations, it will not
  affect the conclusions of this papers since the simulations focus on
  the ability to accurately reconstruct any given intensity and
  spectral index distribution.}.

Two types of datasets were simulated. One was for a VLA single pointing at C-config and L-band
with 16 channels (or spectral windows) between 1 and 2 GHz. The $uv$-coverage was
a series of snapshots \bbf{20 minutes apart, for 4 hours. }
The HPBW of the primary beam at L-band is 30arcmin and therefore covers the
central part of the simulated region of sky. 
The second dataset was for a VLA mosaic at D-config and C-band with 46 pointings (of primary
beams 6arcmin in HPBW) spaced 5 arcmin apart to cover roughly the same patch of sky at
a comparable angular resolution.  
At these simulated angular resolutions( 10arcsec at L-band C-config and 9arcsec at C-band D-config), 
the expected confusion limit is $<0.5\mu Jy$, and the
simulation included only sources brighter than $1\mu Jy$ in order to insulate these tests from 
errors due to main-lobe confusion. 
\bbf{Fig.~\ref{fig:smoothed.model} shows a section of the simulated sky model, smoothed
to 20arcsec (the angular resolution at the low frequency end of the band) in the left panel 
and 10arcsec (corresponding to the angular resolution 
achieved via Multi-Frequency Synthesis (MFS) ) in the right panel. These images
illustrate the level of confusion that subband versus MFS based deconvolution algorithms will 
encounter from the main lobe of the PSF. 
Sidelobe confusion will add to the complexity of the problem in both cases, but more severely
for subband imaging.}
16 channels (or spectral windows) were chosen to span the frequency range of 4-8 GHz,
and the $uv$-coverage corresponds to one pointing snapshot every 6 minutes, tracing
the entire mosaic twice within 8.8 hours. 

A true sky image cube was constructed by evaluating wideband point source components 
from the SKADS list for a set of frequencies that matched those being observed.
All sources were evaluated as delta functions (naturally at pixel centers) using the same
cell size as would be later used during imaging. Specific tests with off-pixel-center
sources were done by using different image cell sizes during simulation and imaging so that
a source at a pixel center during simulation is not at the center during imaging.
Visibilities were simulated per pointing for this image cube, using the WB-A-Projection 
de-gridder \citep{WBAWP} which uses complex antenna aperture illumination functions 
to model primary beams that scale with frequency, have polarization squint
and rotate with time (due to the VLA altitude-azimuth mount). 

Noise was not added to these simulations as our first goal was to characterize numerical limits
purely due to the algorithms and their software implementations. 
Only after all observed trends
and limits are understood will it be instructive to add Gaussian random noise.
Theoretically, pure Gaussian random noise should not change the behaviour of algorithms 
other than increase error in predictable ways and it is important to systematically confirm that 
this is indeed the case in practice. However, numerical noise will be present in these tests at 
the $10^{-7}$ level as most image domain operations use single float precision. All references
to signal-to-noise ratio in this analysis therefore relate to numerical precision noise.


\section{Imaging Algorithms}
The datasets described above were imaged in a variety of ways. In all cases the data
products were continuum intensity images and spectral index maps.
The methods that were tested are several possible combinations of 
standard CLEAN \citep{Hogbom_Clean,Schwab_Cotton_Clean}
for narrow-band imaging, MT-MFS \citep{MSMFS} for wideband imaging, A-Projection \citep{AWProjection}
to account for direction-dependent effects during gridding, and stitched versus joint
mosaics \citep{Cornwell88}. Wideband primary beam correction was applied as appropriate 
using whatever primary beam models were available to the reconstruction algorithms. 
All image reconstruction
runs used the standard major and minor cycle iterative approach \citep{Schwab_Cotton_Clean}
with different combinations of gridding algorithms for the major cycle (prolate spheroid, A-Projection) 
and deconvolution algorithms for the minor cycle (Hogbom Clean, MT-MFS).
The CASA imaging software was used for these simulations and reconstructions as a 
combination of production tasks and custom C++ code and Python scripts. 



\subsection{Cube CLEAN with standard gridding} \label{Sec:algos:single1}

Each frequency channel\footnote{In this paper, a reference to frequency or channel 
implies what is in practice a subband or a collection of data channels that are imaged together
using multi-frequency synthesis with the assumption that the spectrum across its width is
flat or at least ignorable for the desired imaging dynamic ranges.} is reconstructed independently 
with standard narrow-band imaging algorithms (Clean). There is no intrinsic
primary beam correction, but deconvolution is followed by 
post-deconvolution primary-beam correction done per frequency.
All images are then smoothed to the angular resolution of the lowest frequency 
in the observation, spectral models are fitted per pixel to extract spectral indices, 
and channels are collapsed to form a continuum intensity image.

The main advantage of this method is computational simplicity and ease of
parallelization, with each channel and pointing being treated independently.
The disadvantage of this approach is low angular resolution and possible 
sub-optimal imaging fidelity as the
reconstruction process cannot take advantage of the additional constraints
that multi-frequency measurements provide.


\subsection{Cube CLEAN with narrow-band AWProjection}  \label{Sec:algos:single2}
This procedure is the same as Cube, but with projection-based gridding algorithms
applied per channel to account for baseline and time dependent primary beam effects.

AW-Projection uses models of the antenna aperture illumination
functions at different parallactic angles to compute gridding
convolution functions per baseline and timestep\footnote{With an
  altitude-azimuth mount, the antenna aperture rotates w.r.to the sky
  as the antenna tracks a source. The result is a time-variable
  primary beam where the dominant effect is a rotation related to
  parallactic angle.}.  These convolution kernels are constructed to
have conjugate phase structure compared to what exists in the
visibilities, and this eliminates beam squint during gridding.

The main expected differences from standard Cube imaging is in the quality of the
primary beam correction, visible in Stokes V images all the time (beam squint) and
at high dynamic range in the Stokes I image. 


\subsection{Multi-Term MFS with standard gridding}  \label{Sec:algos:single3}
Multi-term multi-frequency synthesis was used to
simultaneously solve for the sky intensity and spectrum, using the
combined wideband $uv$-coverage. With no intrinsic
primary beam corrections, the output spectral Taylor coefficients represent 
the time-averaged product of the sky and 
primary beam spectrum $I^{sky}_{\nu} P_{\nu}$. 
A wideband post-deconvolution correction\footnote{Post-deconvolution wideband PB-correction
is done via [Eq. 14 of REF], a polynomial division carried out in terms of Taylor coefficients. }
of the average primary beam
and its spectrum ($P_{\nu}$) was done at the end to
produce intensity and spectral index maps that represent only the sky.

This method has the advantage of algorithmic simplicity while taking 
advantage of the wideband $uv$-coverage.
\st{and without 
 incuring the computational cost of A-Projection's large gridding convolution kernels.}
The main disadvantage is that the time variability of the antenna primary beam
is ignored, which in the case of squinted and rotating beams can result in 
artifacts and errors in both the intensity and spectrum for sources near the half-power
level.
Also, since the frequency dependence of the primary beam persists through to the
minor cycle modeling stage,  the multi-term reconstruction has to model a spectrum that is
steeper than just that of the sky. More Taylor terms are required, increasing cost and 
low SNR instabilities.


\subsection{Multi-Term MFS with wide-band AWProjection}  \label{Sec:algos:single4}
Multi-term multi-frequency synthesis is used 
along with Wideband A-Projection \cite{WBAWP}, an adaptation of AW-Projection that
uses convolution functions from conjugate frequencies to undo
the frequency dependent effects of the aperture function during gridding
in addition to accounting for beam rotation and squint.
This achieves a clear separation between frequency dependent sky 
and instrument parameters before the sky intensity and spectrum are modeled.
The output spectral Taylor coefficients represent
 $I^{sky}_{\nu} P$ where the effective $P$ is no longer frequency dependent.
In this case, a post-deconvolution division of only the intensity image by an
average primary beam is required. The output spectrum already represents
only that of the sky brightness.
Alternatively, a hybrid of cube imaging with narrow-band A-Projection and MT-MFS can
also be used in which the frequency dependence of the primary beam is removed
in the image domain from the residual image cube before combining the frequency planes
to form the Taylor weighted averages needed for the MT-MFS minor cycle.
These two approaches have different trade-offs in numerical accuracy, 
computational load, memory use and ease of parallelization and a choice between them 
will depend on the particular imaging problem at hand.

This general approach 
has the advantage of accounting for the time and frequency dependence 
of the primary beam during gridding, and clearly separating sky parameters from 
instrumental ones. 
The required number of Taylor coefficients depend only on the
spectrum of the sky, which is usually less steep than that of the primary beam.
\st{The disadvantage of this method is computational cost, although for applications
where its level of numerical accuracy is useful, 
optimized and custom implementations may help. }
With A-Projection, a flat-noise normalization choice can sometimes cause numerical
instabilities around the nulls of the primary beam where the true
sensitivity of the observations is also the lowest -- work is in progress to find a
robust solution to this. Alternate normalization choices will alleviate the problem but
they will increase the degree of approximation that the minor cycle must now handle.


\subsection{Wideband Mosaics for Cube CLEAN and MT-MFS} \label{Sec:algos:mosaic}

In general, a mosaic can be constructed as a weighted average of single pointing images, 
using an average primary beam model as the weighting function. In this discussion,
a combination after deconvolution will be called stitched mosaic, and
a combination before deconvolution will be called a joint mosaic.  A joint mosaic
can also combine the data in the visibility domain by applying
appropriate phase gradients across the gridding convolution functions used by
projection algorithms. 

Several wideband mosaic imaging options exist \citep{WBMOS2014} as various combinations
of imaging with and without DD correction during imaging \st{gridders that use prolate spheroidal functions versus baseline aperture illumination functions},
cube versus multi-term multi-frequency synthesis imaging, and stitched versus joint mosaics.

For Cube CLEAN imaging, joint mosaics were made by combining data in the visibility
domain and applying appropriate phase gradients across gridding convolution functions.
Two methods were compared with the first using 
an azimuthally symmetric primary beam model to construct a single gridding convolution
kernel for all visibilities and the second using full AW-Projection to account for 
PB-rotation and beam squint (per frequency).
The first method has the advantage of computational simplicity compared to full
AW-Projection where convolution functions can potentially be different for every
visibility, but it has the disadvantage of ignoring beam squint and PB-rotation. 
The primary beam model is also not the same as what was used to simulate the data
and this test evaluates the effect of this commonly used simplifying assumption.


For MFS imaging (with multiple Taylor terms), a joint mosaic was computed using 
AW-Projection with its wideband adaptation (to correct for the PB frequency dependence)
along with phase gradients applied to convolution kernels.
An alternate approach is to stitch together sets of
PB-corrected output Taylor coefficient images,using the time-averaged primary beam as
a weighting function, and then recomputing spectral index over the mosaic. 
However, initial tests showed that stitched mosaics (with or without WB-AWProjection) produced
larger errors than joint mosaics \citep{WBMOS2014} and
stitched multi-term MFS mosaics were not included in this analysis.

\section{Metrics to evaluate imaging accuracy}\label{Sec:metrics}

The following metrics were used to evaluate the numerical performance of the different 
algorithms.

\begin{enumerate}
\item Image RMS: The RMS of pixel amplitudes from off-source regions in the image,
or in the case of no source-free regions, the width of a pixel amplitude histogram. 
\item Dynamic Range: Ratio of peak flux to peak artifact level or image rms when
no artifacts are visible.
\item Error distributions (imaging fidelity): The intensity and spectral index maps 
produced by the above algorithms were compared with the known simulated sky
and estimates of error per source were binned into histograms.
For each output image, the simulated sky 
model image was first smoothed to match its angular resolution, and then pixel values
were read off from both images at all the locations of the true source pixels. 
Histograms were plotted for $I/I_{true}$ where deviations from 1.0 indicate
relative flux errors and for $\alpha - \alpha_{true}$ where deviations from 0.0 indicate
relative errors in spectral index. 
All histograms were made with multiple intensity ranges (e.g.Fig.~\ref{fig.lowdr.hist1})
and over different fields of view (e.g. Fig.~\ref{fig.lowdr.hist2}) 
to look for trends.
\end{enumerate}

\section{Single pointing tests and results} \label{Sec:results:single}
The L-band (1-2GHz) C-configuration simulated data were imaged with the algorithms
listed in Sec.~\ref{Sec:algos:single1} through ~\ref{Sec:algos:single4}.

\subsection{Low-dynamic range algorithm comparison}\label{sec.lowdr}

\begin{table*}
  \caption{Intensity and Spectral Index reconstruction accuracy as a function of source intensity}\label{tab:errors1}
  \medskip
  \begin{center}
    \begin{tabular}{cccccc}\hline
      Method  & $I/I_{true}$  & $I/I_{true}$ &  $I/I_{true}$ & $\alpha - \alpha_{true}$ & $\alpha - \alpha_{true}$\\
          Intensity Range          & $>50\mu Jy$  & $8 - 50 \mu Jy$ & $ < 8\mu Jy$ & $>50\mu Jy$  & $8 - 50 \mu Jy$ \\
      \hline
        Cube   & 0.95 $\pm$ 0.05 &  0.9 $\pm$ 0.2 & 0.8$\pm$ 0.3 & -0.15 $\pm$ 0.2 &  -0.1 $\pm$ 0.3 \\
        Cube + AWP  & 1.0 $\pm$ 0.05 &  0.95 $\pm$ 0.2 & 0.9 $\pm$ 0.3 & -0.1 $\pm$ 0.1 &  -0.1 $\pm$ 0.25  \\
        MTMFS & 1.0 $\pm$ 0.02 &  1.0 $\pm$ 0.05 & 1.0 $\pm$ 0.15  & -0.1 $\pm$ 0.2 &  -0.1 $\pm$ 0.2\\
        MTMFS + WB-AWP & 1.0 $\pm$ 0.02 &  1.0 $\pm$ 0.04 & 1.0 $\pm$ 0.15  & 0.05 $\pm$ 0.1 &  0.05 $\pm$ 0.2\\
        \hline
    \end{tabular}\\[5pt]
  \end{center}
\end{table*}

\begin{figure}
\includegraphics[width=3.2in]{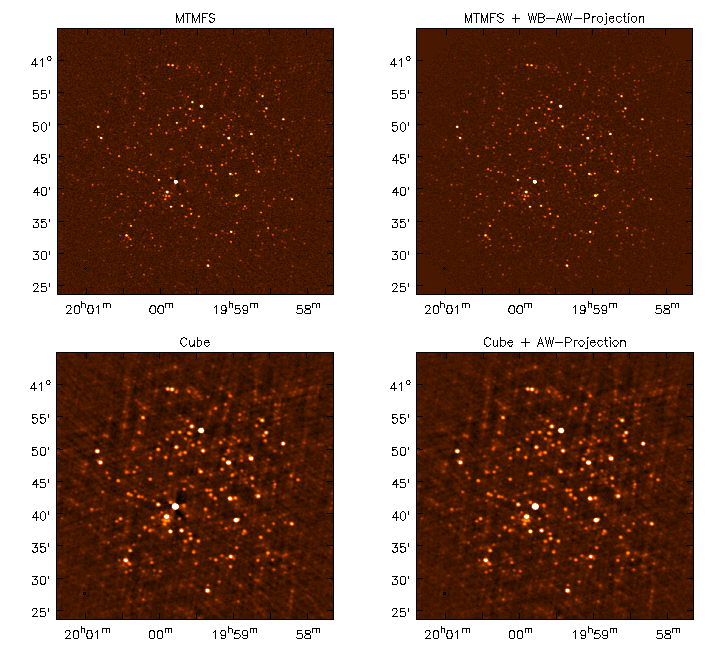}
\caption{Intensity images from the low dynamic range, single pointing tests: 
The top row shows multi-frequency synthesis (MFS) methods that combine all
the data during imaging,  
the bottom row shows cube methods that treat data from each channel independently, 
the left column shows imaging with standard gridding in which PB-corrections done only
as a post-deconvolution division in the image domain, 
and the right column uses A-Projection and its wideband variant
for more accurate PB correction applied before the sky modeling in the minor cycle. 
}
\label{fig.lowdr}
\end{figure}

\begin{figure}
\includegraphics[width=3.2in]{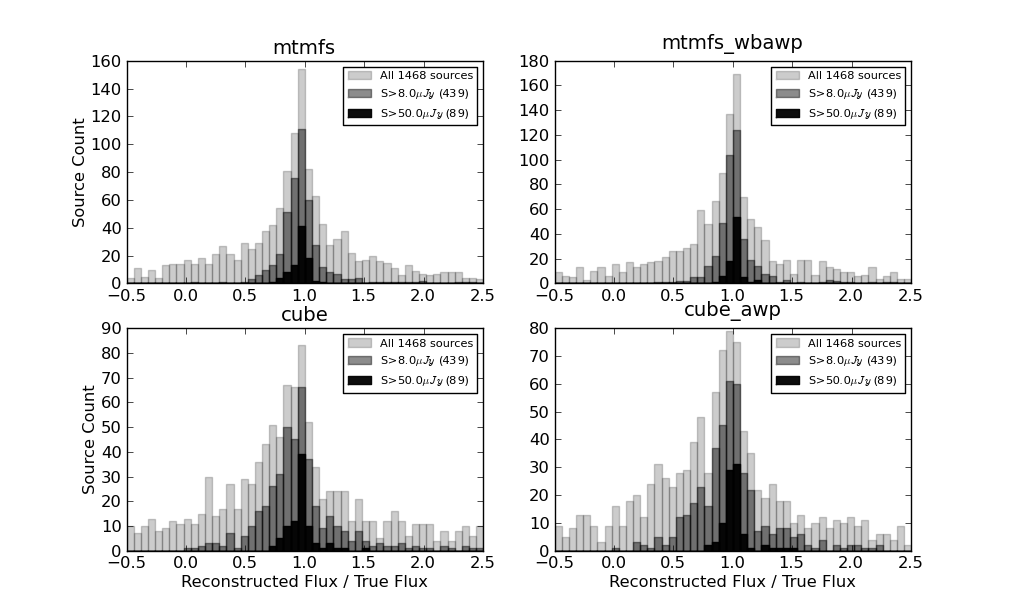}
\includegraphics[width=3.2in]{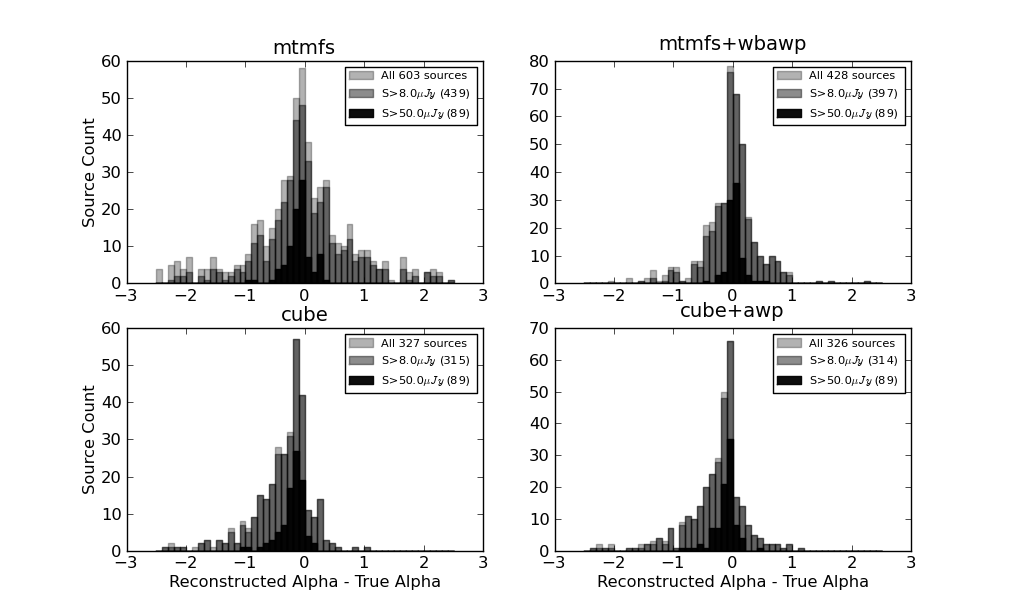}
\caption{Accuracy in Intensity and spectral index as a function of source brightness: 
The top four histograms show $I/I_{true}$ for the four tested methods and the bottom four
histograms show $\alpha - \alpha_{true}$.  Each shade represents sources brighter than a
certain threshold with darker shades representing higher cutoffs. The increased spread for
lighter shades indicates an increase in error for weaker sources. 
}
\label{fig.lowdr.hist1}
\end{figure}

\begin{figure}
\includegraphics[width=3.2in]{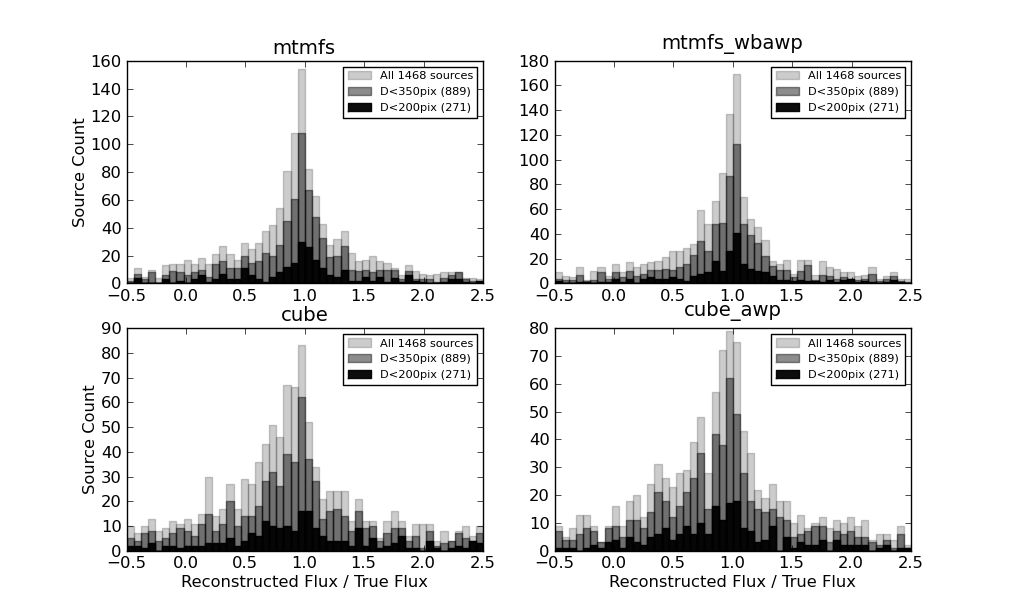}
\includegraphics[width=3.2in]{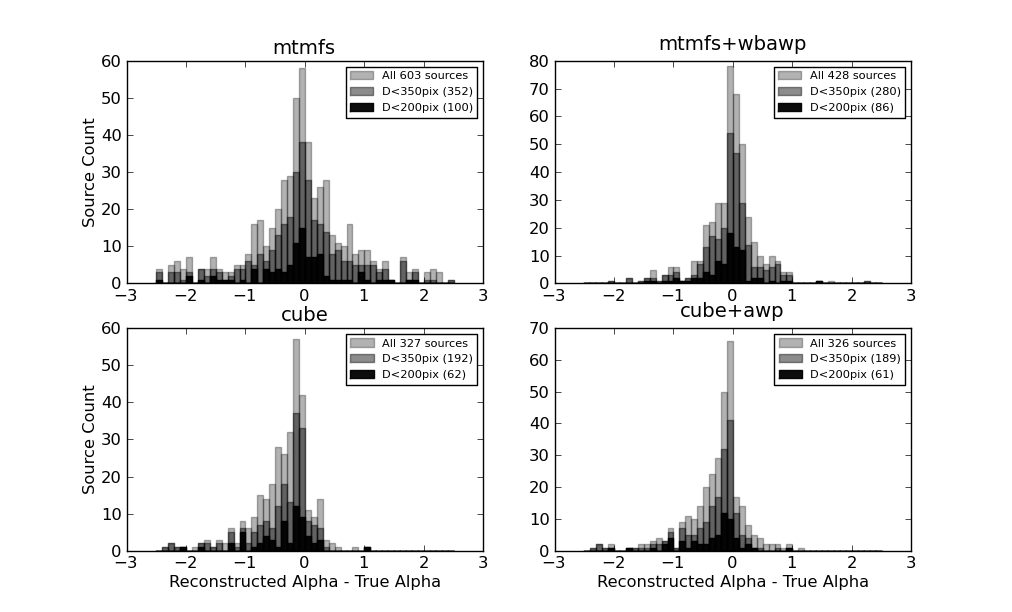}
\caption{Accuracy in Intensity and spectral index as a function of distance from the pointing
center: 
The top four histograms show $I/I_{true}$ for the four tested methods and the bottom four
histograms show $\alpha - \alpha_{true}$.  The shades represent different maximum distances from
the pointing center with darker shades representing the sources closest to the center.
Compared to Fig.\ref{fig.lowdr.hist1}, there are no clear trends of source accuracy as a function
of location in the primary beam. 
 } 
\label{fig.lowdr.hist2}
\end{figure}

The first set of tests used a dataset with visibilities recorded as snapshots every
2 minutes, for 4 hours. The resulting naturally-weighted PSF had a peak sidelobe
level of 0.05 for each channel and 0.02 wideband. 
The brightest source in this dataset was $7mJy$, chosen to demonstrate the
dynamic range limit at which primary beam effects start to show but do not adversely affect 
surrounding sources.  Fig.~\ref{fig.lowdr} shows the intensity images from the four methods, 
with the top row showing multi-frequency synthesis (MFS) methods, 
the bottom row showing cube methods, the left column with PB-corrections done only
post-deconvolution, and the right column with A-Projection for more accurate PB correction. 

The first point to note is the difference in angular resolution between the MFS and
Cube methods. \bbf{The width of the restoring beam was $11arcsec$ with MFS and
$21arcsec$ with the Cube method.}
The second is the presence of many un-deconvolved sources in the Cube methods, 
arising from
weak sources that go undetected in the single channel reconstructions, but show up
only when residual images from all frequencies are stacked.
The third is the Y-shaped artifact pattern around the brightest source in the left column. 
These artifacts are due to PB squint/rotation effects which for the
VLA show up at the $\sim 1000$
dynamic range level at the 0.7 gain level of each primary beam. 
The image rms levels $5e-07 Jy$ show that upto these dynamic ranges, these artifacts 
do not contaminate all surrounding sources. 

The top 4 panels of Fig.~\ref{fig.lowdr.hist1} compares $I/I_{true}$ histograms for 1468 sources 
above $3\mu Jy$ and within the
0.1 gain level of the PB at 1.5GHz. 
The lower 4 panels show $\alpha -\alpha_{true}$ histograms for all sources that met the
threshold for spectral index calculations.
The mean and half-width of each of the resulting distributions (over different intensity
ranges) for these three methods are listed in Table~\ref{tab:errors2}.  

The first point to note is a clear widening of the histogram at
fluxes  $<10 \mu Jy$. Sources brighter than $50\mu Jy$ have errors less than
5\%, sources between $50\mu Jy$ and $8\mu Jy$ show errors at the 10\% level
and sources below $8\mu Jy$ show errors at the 20 to 30\% level with several 
sources more than 50\%.  MFS imaging (top row) shows slightly lower errors compared
to cube imaging especially for weak sources. Cube imaging (bottom row) shows a 
slight bias towards lower brightness for weaker sources and this is most likely to
a residual clean bias effect (see Sec.~\ref{sec:cleanbias}) due to the use of 
single channel PSFs of poorer quality than the MFS PSF.

The trend of brighter sources being more accurate holds
 for spectral indices as well, but the errors degrade much faster
with decreasing source amplitude. Sources brighter than $50\mu Jy$ show errors 
of $\pm 0.15$, but sources between $50\mu Jy$ and $8\mu Jy$ show errors of $\pm 0.5$.
In these tests, the weakest sources did not meet the threshold for 
spectral index calculation and the differences in the number of sources used to compute the
displayed histograms reflects this. The spectral indices are slightly more accurate for the
MT-MFS imaging along with wideband A-Projection and the only trend worth noting is that the
MFS methods produced more sources with usable spectral indices than the cube methods. 
A faint bias towards steepness for low SNR sources is seen for the cube imaging runs and
this could be explained by non-detections at the higher frequencies causing an artificial 
steepening.

Fig.~\ref{fig.lowdr.hist2} shows $I/I_{true}$ and $\alpha -
\alpha_{true}$ histograms with shades representing distance from the
pointing center, out to the 0.2 gain level of the average PB. No
obvious trends were noticed, suggesting that primary beam correction
is equally accurate everywhere within the 0.2 gain level of the PB.
This is expected (and relevant for algorithm validation) as the
  same aperture illumination model was used for simulation and imaging.
  In future tests, errors due to the use of approximate
  models (particularly ones that ignore systematic features)
  during imaging can be evaluated against these results.

\subsection{High-dynamic range algorithm comparison}\label{sec.highdr}

\begin{figure}
\includegraphics[width=3.2in]{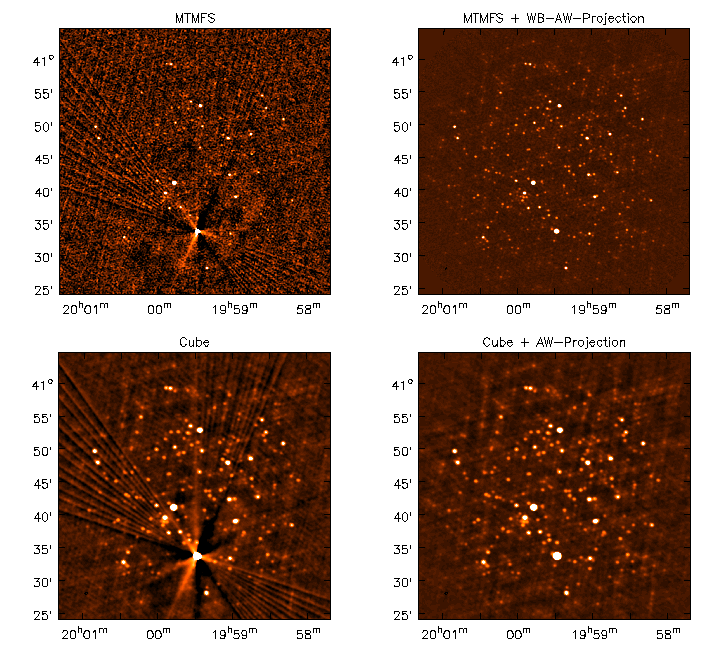}
\caption{Intensity images from the high dynamic range, single pointing tests: 
The top row shows multi-frequency synthesis (MFS) methods that combine all
the data during imaging,  
the bottom row shows cube methods that treat data from each channel independently, 
the left column shows imaging with standard gridding in which PB-corrections done only
as a post-deconvolution division in the image domain, 
and the right column uses A-Projection and its wideband variant
for more accurate PB correction applied before the sky modeling in the minor cycle. 
The effect of a bright $100mJy$ source is clearly visible without and with the use of
A-Projection. 
}
\label{fig.highdr}
\end{figure}

\begin{figure}
\includegraphics[width=3.2in]{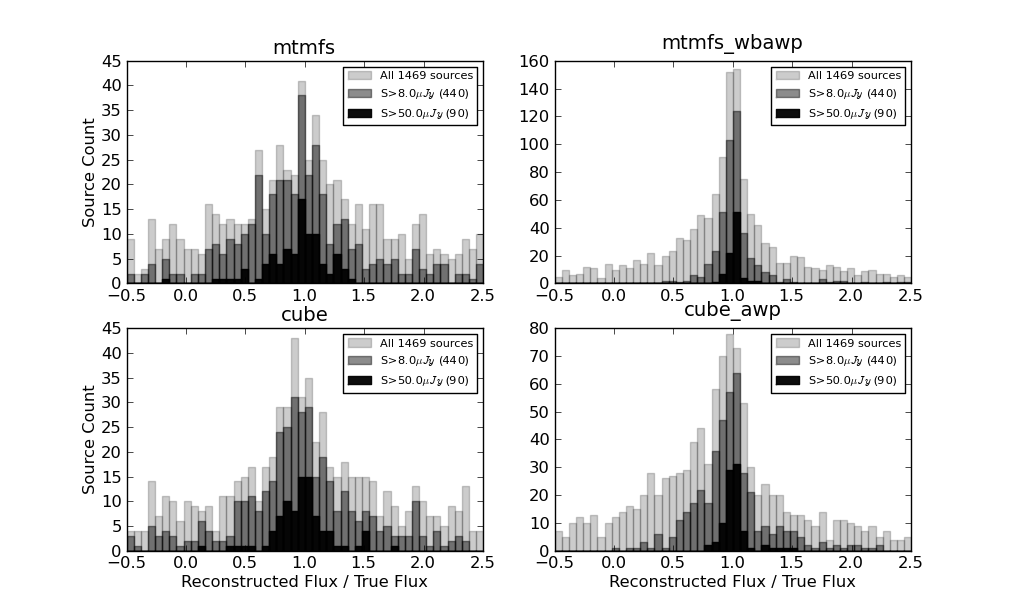}
\includegraphics[width=3.2in]{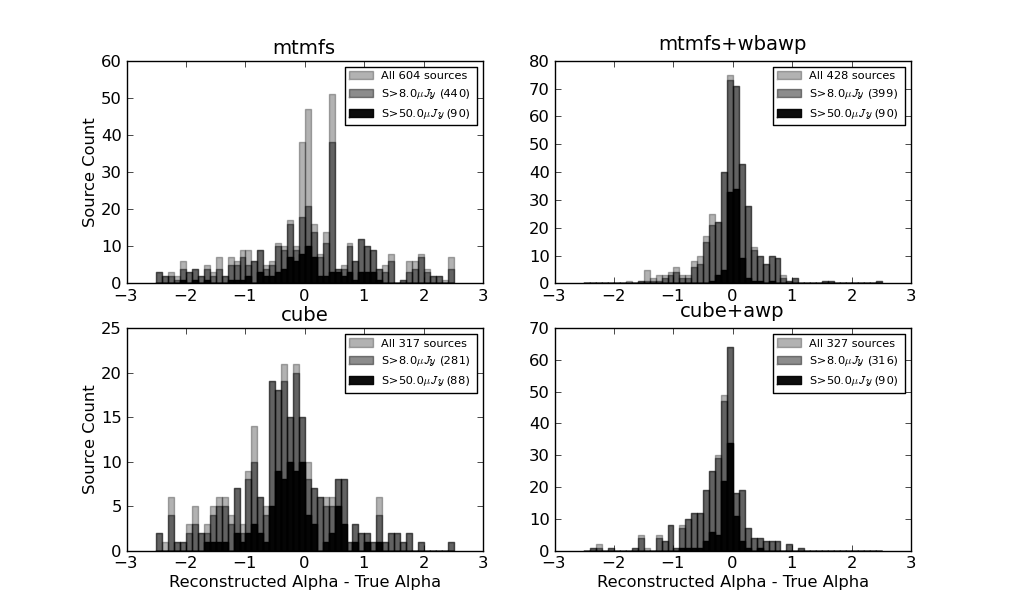}
\caption{Accuracy in Intensity and spectral index as a function of source brightness: 
The top four histograms show $I/I_{true}$ for the four tested methods and the bottom four
histograms show $\alpha - \alpha_{true}$.  Each shade represents sources brighter than a
certain threshold with darker shades representing higher cutoffs. An increased spread for
lighter shades indicates an increase in error for weaker sources. 
}
\label{fig.highdr.hist1}
\end{figure}

The above tests were repeated with one additional $100 mJy$ source added in. 
Fig.~\ref{fig.highdr} shows that the only significant change are artifacts characteristic of uncorrected
primary beam squint in imaging runs that do not do A-Projection. In this case as seen
in Fig.~\ref{fig.highdr.hist1}, the
intensity and spectral index histograms for non A-Projection runs are significantly affected
but after correction are very similar to the low dynamic range example discussed in
section~\ref{sec.lowdr}.

\subsection{Effect of PSF sidelobe level (and data quantity)}\label{sec:sidelobes}

\begin{figure}
\includegraphics[width=1in]{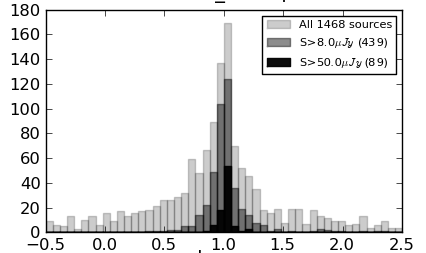}
\includegraphics[width=1in]{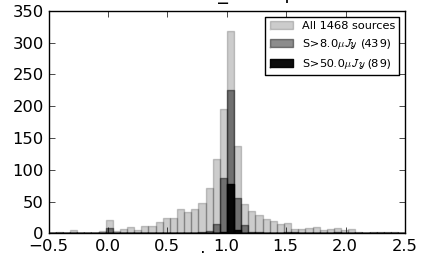}
\includegraphics[width=1in]{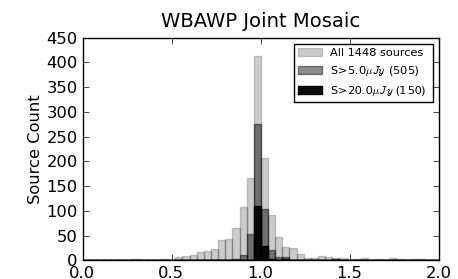}
\caption{Accuracy of intensities as a function of PSF sidelobe level: 
The three panels (from left to right) show histograms of $I/I_{true}$ when PSF sidelobe levels are
0.05, 0.02 and 0.008 due to different amounts of data. 
Shades of grey indicate  different brightness cutoffs. A lower PSF sidelobe 
is clearly advantageous for this imaging run of a very crowded field of point sources even if
source signal-to-noise is not an issue.}
\label{fig.sidelobe}
\end{figure}

Fig.~\ref{fig.sidelobe} compares the accuracies of reconstructed
source intensities for imaging runs involving different $uv$-coverages
and wideband PSF sidelobe levels. The three PSFs in this test had
(from Left to Right) sidelobe levels of 0.05, 0.02 and 0.008.
\st{ The first two PSFs were constructed by selecting \st{subsets of the dataset}
\bbf{different quantities of data} used for
  the low dynamic range simulations described in section
  \ref{sec.lowdr} and the right panel shows the PSF from the joint mosaic 
dataset.}
\bbf{ The first PSF is what was used in the low dynamic range simulations 
described in section  \ref{sec.lowdr}, the second PSF is from a second
set of tests with an increased $uv$ filling factor and the right panel shows 
the PSF from the joint mosaic dataset.}
 These histograms show that as the PSF sidelobe
  levels decrease, sources brighter than $50 \mu Jy$ are always
  reconstructed to within a few percent.  Error on sources in the range $8
  - 50\mu Jy$ improve by more than a factor of 4 (from 20\% to less
  than 5\%) and by more than a factor of 2.5 on
  sources in the range $1 - 8\mu Jy$ (from over $50\%$ to 
  $\sim20\%$).  This behaviour is expected even in the absence of
  noise, simply because of how strongly algorithms like the basic
  Hogbom Clean minor cycle depend on the quality of the PSF when
  deconvolving such crowded fields. The fact that errors change
  significantly even at signal to noise ratios of $\sim 300$ (sources
  at $30 \mu Jy$ and assuming numerical precision noise at $10^{-7}$)
  shows that PSF sidelobe level is still very relevant. 
  Cube imaging is one practical application where this effect may occur
  irrespective of noise level, especially if the narrow-band PSFs have higher PSF sidelobes than
  the joint multi-frequency synthesis PSF.
  This dependence of accuracy on PSF sidelobe level may be weaker for sparser fields where 
  algorithms such as Hogbom Clean are more robust, but this needs verification
  (along with the effects of weighting).



\subsection{Effect of 'clean bias' and use of masks}\label{sec:cleanbias}

\begin{figure}
\includegraphics[width=1in]{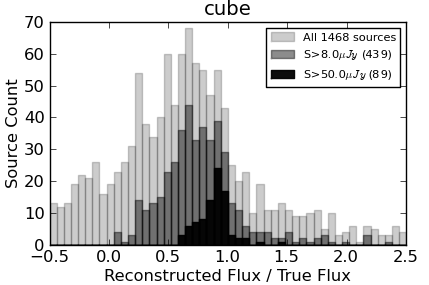}
\includegraphics[width=1in]{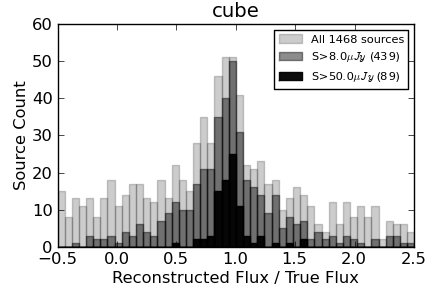}
\includegraphics[width=1in]{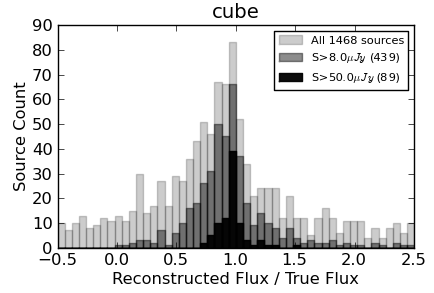}
\caption{Clean Bias: The three panels show  histograms of $I/I_{true}$ for three cube imaging
runs without masks during deconvolution (left), masks around only the 5 brightest sources (middle)
and masks around several tens of sources (right). A systematic downward bias of the 
reconstructed intensity for weaker sources is known as the clean bias. The use of masks or
a PSF with a lower sidelobe level eliminates this problem.}
\label{fig:bias}
\end{figure}

Clean bias, an effect noticed for decades by users of the Clean algorithm, is a 
systematic shift of reconstructed peak intensities to lower than expected values. 
This is usually seen in deep imaging runs with large numbers of closely-spaced weak sources.
We did an imaging test using the relatively sparse $uv$-coverage of 
12 VLA snapshots spread across 4 hours and producing a PSF sidelobe level of 0.15.
With unaided Hogbom Clean in the minor cycle, a clear shift of intensities to lower values 
is seen for weaker sources
(Left panel of Fig.~\ref{fig:bias}). The use of masks or clean boxes to constrain the 
search space alleviates the problem (Middle and right panels of Fig.~\ref{fig:bias}).
A PSF with lower sidelobes (in our case, the full simulated dataset with MFS imaging) 
also prevents this type of flux bias with the Clean algorithm
\bbf {and more importantly it does so without having to invoke complicated masking procedures.}
For example, in our tests, a PSF with sidelobes at 0.13 level showed clean bias
but PSFs with sidelobes at the 0.05 level, did not. 
In cases of high PSF sidelobe levels where clean bias was seen (for example, 
the narrow band PSFs used for Cube imaging), masks 
around the $~$40 brightest sources were sufficient to prevent this bias.

The Clean bias effect can be explained by
considering that the Clean algorithm is an L1-norm basis-pursuit method that 
is optimized for sparse signals that can be described with a minimal number of
basis functions. For astronomical images this implies well-separated
point sources whose properties can be described by single basis functions (one pixel each) 
and whose central peaks are minimally affected by PSF sidelobes from
neighbouring sources. 
With 8000 sources within 0.5 square deg, the imaging problem being discussed
in this paper is certainly not a sparse signal reconstruction, especially with a PSF 
with high sidelobes. The Clean algorithm is therefore error-prone in the low SNR regime.
A systematic lowering of source brightness can be explained by the algorithm
constructing many artificial source components from the sidelobes of real sources.

\subsection{Dealing with un-deconvolved sources}

\begin{figure}
\includegraphics[width=3in]{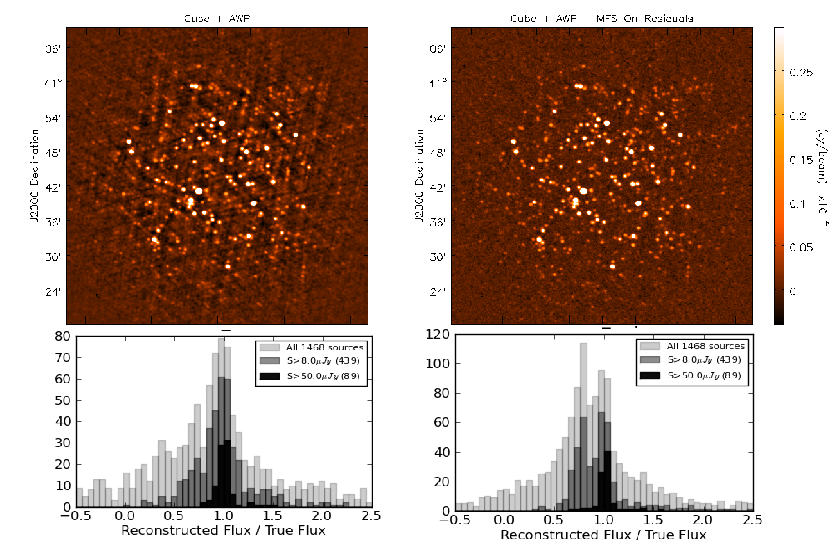}
\caption{Example of a hybrid of MFS and Cube imaging:  The left column shows the intensity
image after just cube imaging and stacking along with the corresponding $I/I_{true}$ histograms.
The right column shows the intensity image and $I/I_{true}$ histograms 
after the second step of continuum imaging on the residuals. }
\label{fig.hybrid}
\end{figure}

With the cube imaging method, the weakest sources are undetected in each narrow band image
but appear as undeconvolved sources when the narrow band restored images are stacked
(seen in the Left panel of Fig.~\ref{fig.hybrid}). A extra deconvolution step can be done with the
continuum residuals, with the assumption that the spectral dependence of weak sources will
not cause artifacts. This is almost always guaranteed to be a valid assumption, since at this
stage the brightest source is at the narrow-band sensitivity limit, and unless the sensitivity
improvement due to bandwidth is more than a factor of $10^3$ across a 2:1 bandwidth, residual 
spectral effects of upto $\alpha=-1$ across a 2:1 bandwidth will not affect the continuum image. 
The right panel of Fig.~\ref{fig.hybrid} shows the resulting image, with a clearly lower 
background noise level and fewer artifacts. The histograms however show a slight degradation in 
accuracy with the hybrid method\footnote{The bimodal structure for weaker sources initally appeared to 
be due to an improper choice of restoring beam size while combining the stacked cube and
MFS continuum model images but we have checked that this is not the case. A numerical or software
error has not yet been ruled out but we are including this example here to illustrate the types of
subtle unexpected effects one may encounter during non-standard usage of 
even straightforward techniques and software. This is also a source or error and uncertainty that is
often ignored in such analyses.}

\subsection{Other tests}
\begin{enumerate}
\item Sources not at pixel centers: Tests were done with sources simulated and imaged using 
image grids of different cell sizes. Thus, what was at the center of a pixel during simulation, is not
at the center during imaging. Slight artifacts were noticed around the brightest sources only beyond
a dynamic range of $10^4$ but no systematic trends were observed.



\bbf{ \item Increased $uv$ filling factor : To control the computing load of these tests, a 
relatively sparse $uv$ coverage was used (VLA snapshots taken 20 minutes apart) in the
tests described above. The results were later compared with a more realistic $uv$-coverage 
constructed from snapshots taken every 2 minutes instead. Errors reduced as expected 
(as seen in Sec.~\ref{sec:sidelobes}) and clean bias was less of an effect but 
there were no changes in trends between algorithm choices. }. 

\item Numerical effects of different convolution function oversampling ratios: 
During gridding with A-Projection, convolution functions are computed at a much finer $uv$ resolution 
compared to the $uv$-grid onto which the data are being gridded. This is to ensure accuracy when
the true $uv$ coordinate is not exactly at the center of an $uv$ grid pixel.  Insufficient oversampling
of the convolution function results in errors similar to using an inaccurate aperture illumination 
function model, and effects such as squint and PB rotation are not corrected for accurately enough. 

\item Differences between software implementations: In practice, different implementations of the 
same algorithm may differ slightly in their numerical details, especially as systems become more
complex. Some such effects are numerical bugs, but some are simply a result of different choices
of normalization and truncation rules.  They produce differences at a level comparable to some 
algorithmic choices and are therefore relevant in such an analysis.

\end{enumerate}

\section{Wideband mosaic tests and results} \label{Sec:results:mosaic}

\begin{figure*}[ht!]
 \includegraphics[width=7in]{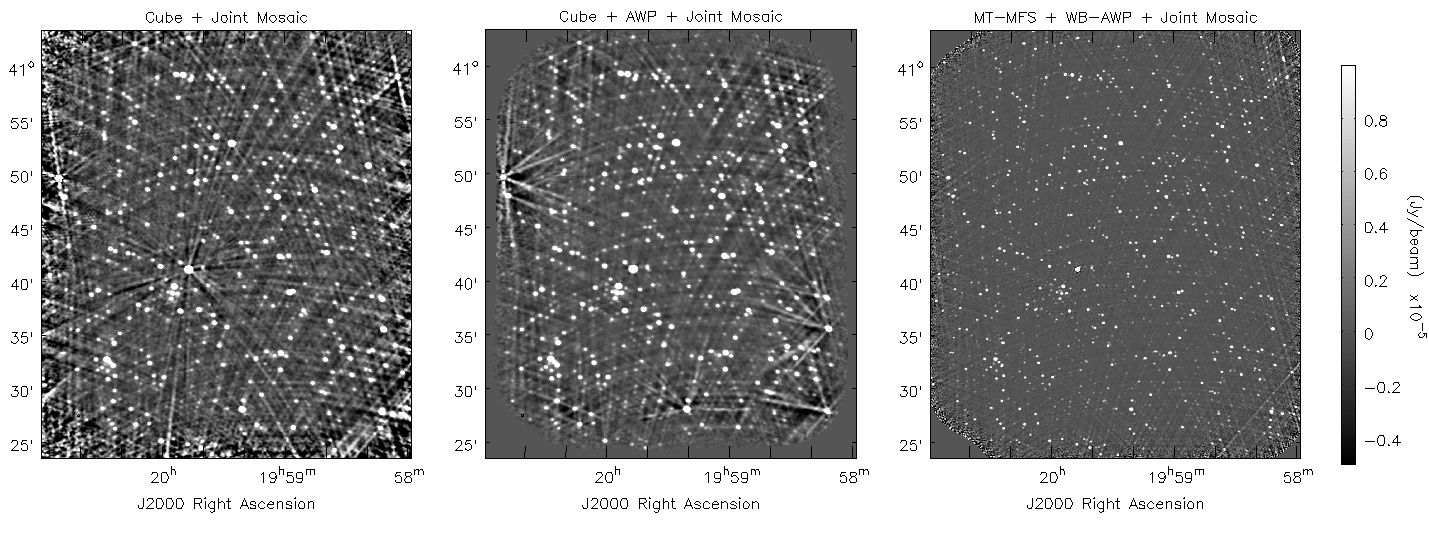}
\caption{Restored continuum intensity images for a wideband mosaic simulation: 
(LEFT) Cube + Joint Mosaic (narrow band imaging using an approximate form of A-Projection in which
the primary beam model is azimuthally symmetric and does not model squint, followed by per channel
primary beam correction before stacking to produce a continuum image and spectral fitting), 
(MIDDLE) Cube + AWP + Joint Mosaic (narrow band imaging using A-Projection that accounts for
azimuthally asymmetric primary beam structure and squint, followed by primary beam correction before 
stacking to produce a continuum image and spectral fitting), 
(RIGHT) MT-MFS + WB-AWP + Joint Mosaic (wideband A-Projection that corrects for the PB frequency
dependence during gridding, combined with the multi-term MFS algorithm with 2 Taylor terms.).
The intensity range shown is $-5\mu Jy$ to $+10\mu Jy$ with the rms in blank regions within the
mosaic field being 4e-08 Jy for the right panel. }
\label{fig.mosimages}
\end{figure*}

\begin{figure*}
\includegraphics[width=2.15in]{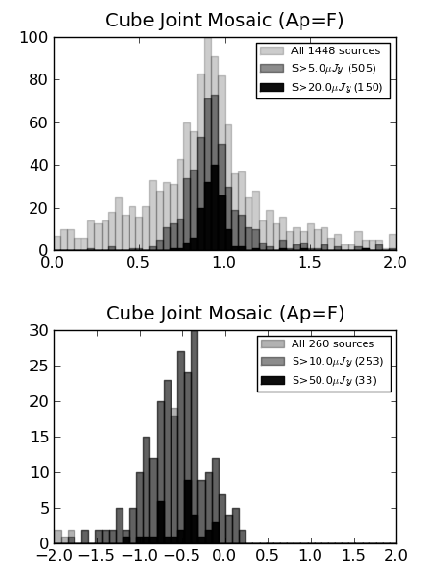}
\includegraphics[width=2.15in]{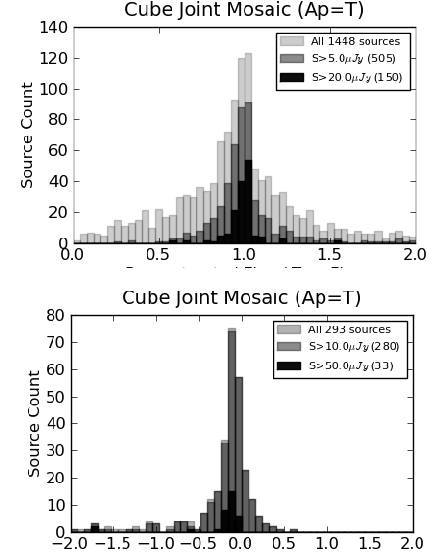}
\includegraphics[width=2.15in]{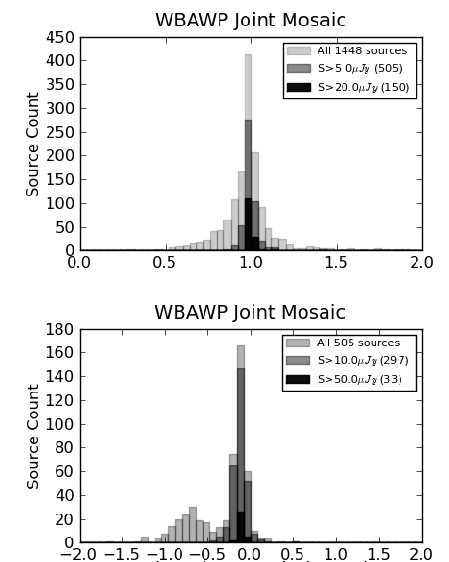}
\caption{Histograms of $I/I_{true}$ and $\alpha - \alpha_{true}$ for the three wideband mosaic
images shown in Fig.~\ref{fig.mosimages}.}
\label{fig.moshist}
\end{figure*}

The C-Band mosaic dataset was imaged using three different methods. 
Fig.~\ref{fig.mosimages} shows the resulting restored continuum mosaic images and 
fig.~\ref{fig.moshist} shows histograms of $I/I_{true}$ and $\alpha - \alpha_{true}$.

The left panel shows the result of cube imaging with 
a joint mosaic using a different primary beam model than what the data were
simulated with and no parallactic angle rotation. 
In addition to the lower angular resolution typical of cube-based methods, 
artifacts around the brightest source are clearly
visible and the intensity and spectral index histograms show considerable scatter.
This test with mismatching PB models between the simulated data and the image
reconstruction algorithm was included to illustrate some of the effects of using 
approximate PBs and algorithmic approximations in practice. The imaging primary beams
were defined only by their radial structure and agree with the time averaged true 
primary beam to within 5\% at the half power point. The brightest source in the
interior of the mosaic field showing artifacts around it is $1 mJy$ in brightness.
This image also   
shows artifacts around sources at the edges of the mosaic, primarily
due to the choice of flat noise image normalization that requires a division by the primary beam
upto a predetermined cutoff level and the inability of the basic Clean algorithm to handle sources
at the very edge of that cutoff level\footnote{An alternate choice of normalization can alleviate this
problem but potentially at the cost of not being able to deconvolve as deeply or accurately 
during the minor cycle and therefore increasing the error in the sky model and amount of computing 
via extra major cycles..}. 

The middle panel shows results from A-Projection (using the appropriate PB model) applied per channel
as part of a joint mosaic. Artifacts around the brightest source have disappeared because the
beam squint and parallactic angle rotation have been accounted for, but as it still is a cube
approach, it has  limited angular resolution. Similar to the left panel, this also shows artifacts around
sources near the edges of the mosaic.   The intensity and spectral index accuracy are however
improved compared to the left panel, and this is simply because of the more accurate wideband 
PB model. 

The right panel shows the imaging results from joint wideband mosaic imaging with multi-term MFS
and wideband A-Projection. The higher angular resolution is immediately apparent and the intensity 
histogram shows a clear improvement over the other two cube methods.
The achieved rms in blank regions within the mosaic field is $0.04 \mu Jy$ 
(in the absence of noise explicitly added to the simulated data).
 The spectral index histogram
also shows a tightening of the errors for bright sources, but a collection of weak sources appear to have a
systematic bias towards steepness. This is clearly an artifact of the algorithm or software. Only
some versions of the imaging run (i.e. some choices of parameters and subsets of the data) show
this effect and no explanation has yet been found for it.
Again, we include this detail in this paper just as a reminder that despite the apparent 
success of the imaging
algorithm as seen from the intensity image (and histogram), numerical surprises may still be 
lurking in the details and tests at the level of detail that are described here are therefore necessary.

The mean and half-width of each of the resulting distributions (over different intensity
ranges) for these three methods are listed in Table~\ref{tab:errors2}.  
Spectral index reconstructions for the weakest
sources $<5\mu Jy$ were not included as all the methods were inaccurate.
These numbers show that cube methods have wider distributions for all intensity ranges.
This is primarily because of weak sources not detected in single channel images but 
appear as confused undeconvolved sources in the continuum image. 
The achieved mean values in both intensity and spectral index
show that accurate handling of the primary beam (via A-Projection)
is required in order to recover the intensity and spectral index to within a few percent and
prevent residual biases, particularly for weak sources. 

\begin{table*}
  \caption{Intensity and Spectral Index reconstruction accuracy as a function of source intensity}\label{tab:errors2}
  \medskip
  \begin{center}
    \begin{tabular}{cccccc}\hline
      Method  & $I/I_{true}$  & $I/I_{true}$ &  $I/I_{true}$ & $\alpha - \alpha_{true}$ & $\alpha - \alpha_{true}$\\
          Intensity Range          & $>20\mu Jy$  & $5 - 20 \mu Jy$ & $ < 5\mu Jy$ & $>50\mu Jy$  & $10 - 50 \mu Jy$ \\
      \hline
        Cube   & 0.9 $\pm$ 0.1 &  0.9 $\pm$ 0.3 & 0.9 $\pm$ 0.5 & -0.5 $\pm$ 0.2 &  -0.6 $\pm$ 0.5 \\
        Cube + AWP  & 1.0 $\pm$ 0.05 &  1.0 $\pm$ 0.2 & 1.0 $\pm$ 0.3 & -0.15 $\pm$ 0.1 &  -0.1 $\pm$ 0.25  \\
        MTMFS + WB-AWP & 1.0 $\pm$ 0.02 &  1.0 $\pm$ 0.04 & 1.0 $\pm$ 0.15  & -0.05 $\pm$ 0.05 &  -0.1 $\pm$ 0.2\\
        \hline
    \end{tabular}\\[5pt]
  \end{center}
\end{table*}

\section{Discussion}\label{Sec:discussion}

The tests described in this paper address the use of wideband data for the
deep imaging of crowded fields of compact  sources 
at a sensitivity close to the confusion limit of the observation. 
The simulations use a realistic source distribution (from which only sources a few times brighter than the
confusion limit were used) and include primary beam effects arising from
azimuthal asymmetry, parallactic angle rotation and frequency scaling. 
\st{The simulated data were noiseless in order to separate numerical and algorithmic effects and limits
from those introduced by noise. }
\bbf {Noise was not added to these simulations as our primary goal for
  this paper was to characterize numerical limits
purely due to the algorithms and their software implementations.}
Imaging results were based on the ability to
apply appropriate primary beam corrections and recover intensities and spectral indices of sources 
out to the 0.2 gain level of the primary beam and down to sensitivities a few times the confusion limit.
Bright sources were introduced to evaluate dynamic range limits with and without corrections for
the time variability of primary beams.

Most observed trends were as expected and we were able to quantify the accuracy with which algorithms
performed \st{in various SNR regimes} \bbf{as a function of source brightness and location in the field of view}. 
There were also  a few surprises that highlight the unpredictability 
of current algorithms in certain situations.

\subsection{Algorithm comparisons}
We compared multiple algorithms and approaches (for example, traditional versus modern)
to understand the limitations of each approach and to decide when the more 
complicated methods really help\footnote{\bbf{Some of } these conclusions are at some level a statement
of the obvious, but given the folklore that often surrounds such analyses and choices 
we feel it is worth re-iterating with clear examples.}.

For a crowded field of compact sources being imaged at a sensitivity
close to the confusion limit, the quality of the PSF matters a great
deal during deconvolution, and multi-frequency synthesis has a clear
advantage over subband based imaging especially for weak sources.  
The Clean bias effect was seen for Cube based methods which required
careful masking to eliminate the effect.
However, MFS-based
wideband algorithms had PSFs with narrower main lobes and lower sidelobes 
and did not suffer from the clean bias
thus making complicated masking procedures unnecessary.

For a mosaic of such a field, a joint imaging approach is
prefered (within reasonable image size limits).  For dynamic ranges
higher than $\sim 10^4$ A-Projection based methods are required to
account for baseline and time-dependent primary beam effects.  Methods
that derive spectral models of the sky while decoupling them from
wideband instrumental effects (MT-MFS with wideband A-Projection)
are capable of producing usable spectral
indices at all SNRs except the lowest SNRs (where all methods fall short).  
The use of WB A-Projection even at
  dynamic ranges lower than $\sim 10^4$ might benefit imaging cases
  where strong sources exist in the outer parts of the PB where
  PB-spectral index is higher.  Finally, simple data parallelization
during the major cycle goes a long way in balancing out the increase
in cost due to the more complex algorithms.

\subsection{Quantifying errors as a function of \st{SNR} \bbf{ source brightness}}
Given the best possible approach 
(multi-term MFS with wideband A-Projection) for the specific problem being studied
(deep widefield imaging of crowded fields at or near the confusion limit), we quantified errors in the
recovered intensity and spectral index as a function of source SNR for both single pointings 
and a joint mosaic. Note that for these noiseless 
simulations there still is numerical noise \bbf{ at the level of $10^{-7} Jy$ } due to the use of single float 
precision in many image domain calculations.

For our single pointing tests (at L-Band), errors in reconstructed intensity were as follows.
Sources brighter than $50\mu Jy$ have errors less than
5\%, sources between $50\mu Jy$ and $8\mu Jy$ show errors at the 10\% level
and sources below $8\mu Jy$ show errors at the 20 to 30\% level with several 
sources more than 50\%.
Errors in reconstructed spectral index were as follows.
Sources brighter than $50\mu Jy$ show errors 
of $\pm 0.15$, but sources between $50\mu Jy$ and $8\mu Jy$ show errors of $\pm 0.5$.
In these tests, the weakest sources did not meet the threshold for 
spectral index calculation (for both cube and multi-term MFS methods). 

For such crowded fields, accuracy also depends on the quality of the PSF even if 
source SNR is not a problem. We showed that as PSF sidelobe levels decreased
\bbf{ from 5\% to 2\% and then 0.8\% of the peak} (by choosing different subsets of the data) 
sources brighter than $50 \mu Jy$
are always reconstructed to within a few percent, sources between 8 and 50 $\mu Jy$ improve from
20\% errors to less than 5\%, and sources between $1 \mu Hy$ and $8 \mu Jy$ improve from over 50\%
errors to about 20\%.  

For our joint mosaic tests (at C-Band), errors in reconstructed intensity were as follows.
Sources brighter than $20\mu Jy$ had 2\% errors, sources between $20\mu Jy$ and $5\mu Jy$
show errors of 4\% and sources below $8\mu Jy$ have about $15\%$ errors.
Spectral index errors were $\pm0.05$ for sources brighter than $50\mu Jy$ but sources
between $50\mu Jy$ and $10\mu Jy$ had errors upto $\pm 0.2$ with both cube and MFS 
methods showing a slight systematic bias of $0.1$ towards flatness. 
\bbf{ The use of an approximate primary beam model (that agreed with the time-averaged ideal beam to within
5\% at the half-power level) caused an increase in the error at
all brightness levels as well as spurious biases in the reconstructed intensity and spectral index. }
The difference between the scale of the errors between the single pointing and mosaic tests
relate to the amount of data used in the simulation.

\subsection{Example of simulation complexity}
This study is a demonstration of the level of detail at which such simulations must be 
carried out in order to begin to be useful as an accurate predictor of reality. 

There are many sources of error during image reconstruction and the systematic separation of
various contributing factors and their eventual combination is crucial to building a 
complete picture and truly understanding the reasons behind observed effects. Conclusions 
derived from approximations done in isolation must be treated with caution and 
trends observed in real data must be reproduced when applicable. In a first step we include 
artifacts typical of various instrumental effects (primary beams, PSF sidelobe levels)
and demonstrate clean bias and show how to eliminate it.
In fact, even the simulations done in this paper
require the addition of several more effects to become accurate predictors of reality.
For example, this paper quantifies algorithmic limits in the situation of no noise, point sources located at
pixel centers, and no differences between the primary beam models used for simulation versus 
those used during imaging (except for one of the wideband mosaic tests). 
These errors are thereforem a lower limit on what one could expect in
reality.  Effects such as the clean bias were reproduced clearly and its cause 
and solution understood.
Future enhancements (even just for Stokes I imaging)  should include noise as well as residual
calibration errors (some of which may masquerade as effects needing baseline based calibration),
the use of inaccurate primary beam models during imaging, the presence of extended emission
in addition to point sources, etc. 
Such questions are listed in detail in Sec.~\ref{Sec:openqns}.


\subsection{Accuracy vs computational cost}

Sometimes, algorithmic choices and achievable numerical accuracy depend on the
type of available computing resources. This section revisits the various algorithmic
options in the context of numerical accuracy versus computational cost.


Cube imaging methods are the easiest to parallelize, with both data and images being
partitioned across frequency for the entire iterative imaging process. There is minimal
need for special-purpose software for such a setup. Imaging accuracy is limited to that
offered by the $uv$ coverage per channel, deconvolution depth is limited to the single
channel sensitivity, and the resolution at which spectral structure can be calculated is 
limited to that of the lowest frequency. There is no dependence on any particular
spectral model which makes this approach very flexible in its reconstruction of 
spectral structure.

Multi-frequency synthesis is demonstrably superior for continuum imaging due to its
increased angular resolution, imaging sensitivity and fidelity, especially for crowded fields
with thousands of compact sources.
The multi-term MFS algorithm is useful to compute in-band spectral indices along with intensity
but the cost of both the major and minor cycle increase with the order
of the polynomials used.  Also, the accuracy of the spectral indices
depends on the source SNR\footnote{Note that the dependence of the
  spectral index accuracy on SNR is fundamental and not limited to a
  particular algorithm (e.g. MT-MFS) or procedure (e.g. Cube imaging)
  used to measure spectral indices.} and the choice of the order of
the polynomial. Work is in progress to test an approach where the
number of Taylor terms is SNR dependent.  For multi-frequency
synthesis, only the major cycle can be easily parallelized, with a
gather step performed before the joint deconvolution step. The
prefered partition axis when the WB-AWP algorithm is used is time
because of the use of aperture illumination functions from conjugate
frequencies during gridding.  If narrow-band A-Projection is used to
form a cube before primary beam correction and the formation of Taylor
weighted averages in the image domain, the partition axis of choice
for the major cycle would be frequency.

Projection algorithms are significantly more expensive than the more
standard method of using prolate spheroidal functions during gridding,
mainly because of the support size of the convolution kernels and the
overhead of computing such functions for potentially every visibility.
However, the more useful metric is the {\it total} runtime for imaging and the
extra cost of using projection algorithms can be offset by a 
comparable reduction in the runtime due to its numerical advantages.
For example, WB A-Projection increases the computing load for imaging but
decreases the computing load and memory footprint of the MT-MFS
setup which will need fewer terms to fit the spectral structure since the
primary beam spectrum has been eliminated. 
Also, in practice the roughly 10 fold increase in
computation due to the use of A-Projection compared to the standard
gridder is readily absorbed by simple data parallelization during the
major cycle.  In addition, approximations can always be made
(identical antennas, coarser sampling of the aperture illumination
function to reduce the size of the convolution functions, etc) but
effects of such approximations are visible beyond the $\sim 10^5$
dynamic range level.

Another axis along which parallelization is relatively easy is pointing. However, our tests show
that the numerical differences between stitched vs joint mosaics are large enough that
(for crowded fields) joint mosaics are always preferred. 
Another basic factor is the use of single vs double precision calculations during imaging
and deconvolution. Currently (in CASA), all intermediate and output images use single
precision, which is not the best option for dynamic ranges $\ge 10^7$.

In summary, an appropriate choice of algorithm depends  (as always) on the desired 
angular resolution, imaging sensitivity (and fidelity), dynamic range, data volume and
available computing resources.

\subsection{Open questions}\label{Sec:openqns}
The simulations and tests described in this paper 
demonstrate several ways in which imaging accuracy 
can be sub-optimal even for the simple situation of point sources imaged using both traditional 
trusted techniques as well as newer ones.
A vast number of open questions and details remain, and a truly accurate picture can be derived 
only after these avenues are explored carefully and quantified to provide trends and
usage guidelines to astronomers.

Work is in progress on several of these fronts, and results will be presented in subsequent
papers.

\begin{enumerate}
\item Is it better to trade integration time at a single frequency band for shorter samples taken across 
a wider frequency range ? For example for the VLA, simulations have shown that comparable 
imaging sensitivities and far more accurate spectral indices are recovered when an observation
spans multiple bands (L-Band and C-Band for example) compared with the entire time spent at
only one band. 
\item What is the best algorithm for emission consisting of extended structure as well as compact
emission ? Algorithms like Multiscale CLEAN are usable but relatively more expensive in terms of computing
  and memory footprint load and require considerable human 
input.  Several newer methods have been shown to produce superior results on their own but
do not currently have production-quality optimized implementations that one can use.
\item Does the addition of noise and residual calibration errors change any of the above 
conclusions ? Theoretically, one would not expect
the addition of Gaussian random noise to change any results but it will be instructive to understand 
how robust these algorithms are to various noise levels \st{and to see if any new systematic trends 
appear (especially in the light of recent anecdotal reports of a flatness bias in spectral indices in 
simulations that contain noise but which ignore important primary beam effects.)}.
Residual calibration errors on the
other hand might cause changes that must be quantified (e.g., see \cite{ADATTA2009}). 
To assess how well such effects can 
be corrected, traditional self-calibration techniques must be compared with more flexible 
methods such as direction dependent calibration schemes and peeling (for example, SAGE), 
especially to test their effects on the accuracy of the reconstructed sources.
\item How does baseline based averaging affect the achievable accuracy in the reconstructed
intensity and spectral index ?
A popular mode of data compression is to average time-contiguous visibilities
that will all fall on the same $uv$ grid cell during gridding. One concern with such an approach is
whether it would prevent the handling of time variability of directional dependent instrumental effects or not.
A simple test that achieved a data size reduction of 20\% for one of the test datasets showed no 
noticeable effect with A-Projection imaging out to twice the HPBW of the PB. Additional tests must be done
with more practical data compression ratios. 
\item How effective is the standard P(D) analyses in predicting source counts below confusion limits ?
Simulations similar to those described in this paper with sources weaker than $1\mu Jy$ (or an observation 
with a larger angular resolution) can be used to test the effect of main-lobe confusion
for the simulated source count distribution and the accuracy of P(D) analyses on such an image. 
\item What type of software implementation and parallelization strategy is the most appropriate for
a particular type of survey ?  In the past few years several new modern imagers have begun to become
available and it would be instructive to repeat an imaging test with different algorithms
and implementations to evaluate and quantify differences that arise simply from different software 
implementations and subtle numerical and algorithmic choices within it. 
For example, \cite{WBMOS2014} shows examples of the numerical differences one can achieve 
simply by using PB models of different kinds and different algorithmic and software implementations
for wideband mosaic imaging.
\item How do these results extend to full polarization imaging and Faraday rotation synthesis? 
Simulations with polarization dependent primary beams and full-stokes imaging (with and without
A-Projection) can quantify polarization imaging limits and identify appropriate imaging strategies.
Work is in advanced stages  \citep{PJ_THESIS}.
to analyse this and produce the required algorithms and software to do such imaging.
\item How accurate do primary beam models need to be for use within A-Projection ? 
Simulations with controlled differences in actual aperture illumination functions can give a useful idea of 
how much variation can be left unmodeled during imaging. For example, it is easy to produce
a spurious bias in spectral index simply by using a PB model whose shape is slightly different from 
what is present in the data.  Work is in progress \citep{ALMAPB2016} to quantity
imaging errors for ALMA when (not so) subtle differences between antenna structures and illumination
patterns are ignored at different levels of approximation and for the VLA  \citep{PJ_THESIS}
to carefully model primary 
beams from holography data and use them during image reconstruction.

\end{enumerate}

\subsection{Conclusions}

These tests probe the limits of commonly used interferometric imaging algorithms 
in the context of crowded fields of compact sources being imaged at a few times the 
confusion limit and $>10$ times the (numerical) noise level.
\st{Such a study is of relevance right now for existing telescopes that are awarding time for large
deep surveys (for example VLASS with VLA, EMU/POSSM with ASKAP, MIGHTEE with MeerKAT), 
as well as for the dish component of SKA that aims to routinely process multi-TB 
datasets at microwave frequencies to produce accurate images at $> 10^7$ dynamic 
ranges and $100 nJy$ flux levels with high levels of fidelity.  }


\paragraph{PSF quality with crowded fields : }
In this regime, the quality of the PSF is of considerable importance even at signal-to-noise
ratios of $>100$ simply because of the limitations of Clean based deconvolution
algorithms in crowded fields. A PSF sidelobe of $<1.0\%$ 
 to achieve errors of $<10\%$ in intensity and $<0.2$ in spectral index across a 1-2GHz band 
for low brightness sources near the confusion limit. 
Since PSFs from the joint imaging of data (MFS, for example) typically have
lower sidelobes compared to PSFs from partitioned pieces of data
(Cube, for example) the former are prefered algorithmic choice in this
regime.
%
%

\bbf{ \paragraph{Clean bias and the need for masks :}
Clean bias is a relevant effect in crowded fields when the PSF sidelobe level 
is high ($>10\%$ of the peak in our tests). In our tests, Cube 
methods showed Clean bias and detailed masks were needed to assist the 
deconvolution algorithms and eliminate the effect. 
However, MFS-based
wideband algorithms had lower PSF sidelobes and did not suffer from the clean bias
thus making complicated masking procedures unnecessary. 
}

\paragraph{Sparse fields are easier :}
For shallow surveys where the emission above thermal noise limits
  fills the sky sparsely, we did not find any statistically
  significant difference between algorithms that partition the data
  and those that don't (like MFS).  Surveys that require imaging at
  the native resolution of the data/telescope -- particularly where
  detection and reconstruction of extended emission is important --
  will still need to use MFS.  Computing resources will be another
  discriminator in choosing algorithms in the shallow regime with
  algorithms that don't require data partitioning (like MFS) requiring
  fewer resources than algorithms that do require partitioning.

\paragraph{High dynamic range : }
For wide-field imaging (DR $>10^4$), primary beam correction
  methods need to include details such as its time, frequency and
  polarization dependence to both reconstruct the bright source
  accurately and to eliminate artifacts that may contaminate
  surrounding weaker sources. Our investigation shows that accounting
  for azimuthal asymmetry of the beams, rotation or pointing jitter as
  a function of time, scaling with frequency and polarization beam
  squint due to off-axis feed locations is certainly required.

\bbf{ \paragraph{Errors as a function of algorithm, brightness and location : }
With all algorithm combinations we tested on these noiseless simulations, 
errors on the reconstructed intensity increased from less than 5\% for 
bright sources to $>$20\% for the weakest micro-Jy level sources.
Errors on the spectral index varied from $\pm 0.15$ for bright sources
to upto $\pm 0.5$ for
the weakest sources and degraded faster than the errors on intensity.
Cube based algorithms consistently had slightly larger errors at all intensities 
compared to MFS based methods.
For wide-fields of view, there were no noticeable trends in the errors as
a function of distance from the pointing center as long as an accurate
wideband primary beam model was used.
However, as demonstrated via one of the wideband mosaicing tests, 
the use of an approximate primary beam model 
(that agreed with the time-averaged ideal beam to within
5\% at the half-power level)
can cause spurious biases in both the intensity and spectral index. 
}

\paragraph{Computational load : }
The imaging performance of joint image reconstruction methods is
  fundamentally superior to those that work with partitioned data and
  combining number of reconstructed images each from a fraction of the
  available data.  Hybrid implementations that take advantage
  of the ease of parallelization of partitioned methods where possible
  may be useful, but require careful
  analysis of its final imaging performance \bbf{ and numerical accuracy}.  
\st{  Our analysis shows that hybrid
  methods are certainly not computing-resource efficient.}
\bbf{As a general rule, in order to benefit maximally from advanced 
algorithms, instrumental and sky effects should be separated as early in the 
  image reconstruction process as possible.}

\paragraph{Simulation complexity : }
Work presented here also gives a 
  lower limit on the level of detail at which simulations for future
  surveys and telescopes must be undertaken.  
  It is important to systematically build up the complexity of the simulation
  in a way that enables one to efficiently pinpoint the reasons behind observed 
  feature or trends \bbf{ before moving on and adding more complicating details}.
  Given the wide range of
  observation types and analysis methods, simulations must be specific
  to the parameters of each survey and must be as complete as possible
  in their inclusion of instrumental effects and observing modes. In
  particular, effects of time, frequency and polarization dependence
  of the antenna primary beams, effects of long baselines and wide
  fractional bandwidths, and correct sampling of the coherence field
  in time and frequency to realistically reflect both the data volume
  and the filling factor in the uv-plane.  Clear accuracy requirements 
  are also necessary for each survey in order to choose
  the optimal analysis procedure based on scientific estimates.



\begin{acknowledgements}
We wish to thank the various NRAO staff members for useful discussions at various
stages of this project. 
We wish to thank the Common Astronomy Software Applications (CASA) Group for the 
use of their imaging libraries in our simulation and imaging scripts. 
\end{acknowledgements}

\bibliographystyle{apj}

\end{document}